# User Behavior Assessment Towards Biometric Facial Recognition System: A SEM-Neural Network Approach

Sheikh Muhamad Hizam[1], Waqas Ahmed[1(✉)], Muhammad Fahad[2], Habiba Akter[1], Ilham Sentosa[1], Jawad Ali[3]

[1] UniKL Business School (UBIS), Universiti Kuala Lumpur, Kuala Lumpur 50300, Malaysia
[2] COMSATS University Islamabad, Attock Campus, Pakistan
[3] Malaysian Institute of Information Technology (MIIT), Universiti Kuala Lumpur, Kuala Lumpur, Malaysia
`waqas.ahmed@s.unikl.edu.my`

**Abstract.** A smart home is grounded on the sensors that endure automation, safety, and structural integration. The security mechanism in digital setup possesses vibrant prominence and the biometric facial recognition system is novel addition to accrue the smart home features. Understanding the implementation of such technology is the outcome of user behavior modeling. However, there is the paucity of empirical research that explains the role of cognitive, functional, and social aspects of end-user's acceptance behavior towards biometric facial recognition systems at homes. Therefore, a causal research survey was conducted to comprehend the behavioral intention towards the use of a biometric facial recognition system. Technology Acceptance Model (TAM) was implied with Perceived System Quality (PSQ) and Social Influence (SI) to hypothesize the conceptual framework. Data was collected from 475 respondents through online questionnaires. Structural Equation Modeling (SEM) and Artificial Neural Network (ANN) were employed to analyze the surveyed data. The results showed that all the variables of the proposed framework significantly affected the behavioral intention to use the system. The PSQ appeared as the noteworthy predictor towards biometric facial recognition system usability through regression and sensitivity analyses. A multi-analytical approach towards understanding the technology user behavior will support the efficient decision-making process in Human-centric computing.

## 1 Introduction

As technology is permeating around human life, the living and interacting scenarios have become smart. The smartness belongs to intelligent systems such as the Internet of Things (IoT), Artificial Intelligence, sensors, etc. in the shape of smart homes, smart offices, smart streets, and smart cities. The purpose of such an intelligent system is to develop sustainability in the knowledge economies for all stakeholders to make life more manageable, secure, and efficient. The government uses this technology to solve the issues of governance, the denizens enjoy the applications of innovation in their life and businesses digitally manage their functions and support both the govt and citizens with such technological solutions. In this scenario, using the technology for security plate-form entails the dynamic implication in surveillance, observing, and scrutinizing the subject activities at homes, workplaces, and on streets. However, advances in technology compel the up-gradation of security and privacy of users and their activities at home and offices. Taking the security concern and technology use, the conventional ways were based on the passcodes, keys, cards, or sensors devices. The novel addition to smart security features is the biometric facial recognition system. This system works through an algorithm by identifying the human face through nodes and matching them with already available records in databases to proceed for the desired action like opening the door or to alert authorities. The main purpose of this system is to enhance the security and keep the record of activities without human interference or manned involvement at homes, institutions, and workplaces. The efficacy of biometric facial recognition involves high accuracy even in the crowds with the non-contact nature. This smart security system is worked in criminal identification, tracking attendances, airport screening, defense service, online payment, and banking services.

Towards the home security purpose, biometric facial recognition is believed to be the most efficient system as compared to other security systems like fingerprints, retina scanning, or passcodes, etc. The efficacy of this system at homes lies in no-contact nature technology. This system provides accurate control of any manned activity being passed in the home. The technology works through sensing the humanoid entrance to the home, scanning the face of a person who may be a friend or family member via camera, then matching the facial nodes with database images, and open the door/gate upon identical results[1]. Then the homeowner gets information through mobile applications about the activity. The overall process fortifies the smart home scenario by the IoT mechanism. The foremost advantages of such a system are accuracy, convenience, and hygienic way of entrance. This feature of the smart home affords the automation and smart integration towards other activities such as control of electric appliances, electronic gadgets, and synchronizing with home temperature sensors, etc. Biometric facial recognition based smart security system is debated by the quality features of its performance [2] such as accuracy level in day and night mode, customized confidence, and differentiating the original human face versus masks or pictures, etc.

The core advantage of innovation relies on its usability level. The interaction of end-users with novel technology narrates through its features, functionality, and user convenience [3]. Besides the technical aspects and usability features of digital tools, the cognitive process and social aspects also engage the willingness behavior to use the technology. Identifying the key elements that affect the user's behavior towards technology use is considered as the primary goal of Human-centric computing (HCC) study. HCC is based on how human interacts and make use of computing technologies in their daily life through personal and professional ways. This phenomenon entails the user behavioral insight which is profoundly explained through Human-Computer Interaction (HCI). By taking biometric facial recognition technology in the HCI context, this study aims to comprehend the usability factors by seeing the system quality features, psychological aspects, and societal elements. Being the novel technology, the literature on usability factors in facial recognition is limited to few studies, and researches specific to home-users acceptance towards facial recognition are appeared to be scarce. By looking into the HCI literature, the facial recognition payment system in China was assessed by users' cognitive capabilities and system characteristics through SEM quantitative analysis[4]. In another study, the usability intention of the hotel facial recognition system was determined through the emotional sense of guests with system features[5]. In the video surveillance acceptance study, Technology Acceptance Model (TAM) was undertaken to assess the citizen behavioral determinants[6]. Towards biometric identification devices, the behavioral intention was measured by personal and technical factors through structural equation modeling (SEM)[7].

As the literature on facial recognition is unable to encapsulate the behavioral aspects of home technology users. There is merely one study on the acceptance of facial recognition system with the solitary postulation of the reliability analysis [8] instead of SEM towards behavior assessment with the small sample size that is insufficient to draw the framework for behavioral modeling in innovation adoption [9][10]. Besides, there is a paucity of study explaining the role of the contextual and personal factor in determining the antecedents of biometric facial recognition through detailed behavioral analysis. This research intended to classify the behavioral aspects through the TAM model, which explains the human cognitive features in acceptance decision, along with social influence and perceived system quality. A novel hybrid quantitative analysis technique i.e., SEM-Neural undertook to make study inferences dynamic and validated towards the prediction of the vital factors. Artificial Neural Network (ANN) or simply Neural is a highly valid and progressive analysis method uniting SEM towards end-users behavioral prediction in digital technology acceptance[11][12]. The research will bring an understanding of smart home features integration from a security point of view i.e., biometric facial recognition system. This research had conducted secondary data collection from previous literature of HCI and facial recognition system to develop the conceptual framework. Primary data assembly was managed from potential intended users of facial recognition systems at homes to conduct the SEM-Neural Analysis.

## 2 Literature Review

Understanding the technology user behavior is an elementary key to infuse the innovation and to make the implementation process smooth and profound. There are various theories pertaining the types of characteristics to comprehend the user behavior such as the Theory of Planned Behavior (TPB), Technology Acceptance Model (TAM), TAM 2, Unified Theory of Acceptance and Use of Technology (UTAUT), Diffusion of Innovation (DOI), Technology-Organizational-Environment framework (TOE), etc. The drive of these models is to determine the adoption factors by keeping the human factor above contextual scenarios. By keeping the view of literature on IS theories and face recognition studies, the TAM model was considered as a base theory with perceived system quality and social influence. TAM is the most validated theory in the HCI context to predict the technology use behavior from primary school children to fighter jet pilots. It stations on two cognitive capability elements i.e., perceived usefulness (PU) and perceived ease of use (PEOU). This duo of users' perceptions of system performance significance i.e., PU and system convenience of use i.e., PEOU builds the positive and negative attitude towards usability. Eventually, this attitude strengthens the behavioral intention to use or resist adopting the technology. In recent studies, PU and PEOU directly correlated the behavior intention (INT) without expressing the attitudinal countenance [13]. In the tourism sector, the facial recognition technology significance was hypothesized with performance expectancy which is the parallel term of PU towards INT of system use [5]. Likewise, effort expectancy, which is an advanced term of the TAM model's PEOU factor, was assessed in payment facial recognition and found a vibrant influencer towards behavioral intention[4]. By analyzing the relationship between PU and PEOU towards willingness behavior to use a biometric facial recognition system explores the perception of the usefulness of technology and the level of convenience in the system usage respectively. Consequently, in the home security scenario, PU and PEOU of biometric facial recognition systems can positively influence the willingness to use the system.

System characteristics, structures, and dynamics of functionality usually position in the service adoption and term as the quality landscapes from the HCI context. Using the biometric facial recognition system at home, nodes towards the privacy and functionality of technology. This system has less influence over the users in terms of word of mouth and other means of dissemination due to its novelty in the market. System quality in acceptance of video surveillance frolicked a significant role in behavioral intention modeling [6]. Quality elements for IoT services consist of privacy, functionality, efficiency, and tangibility [3]. Perceived system quality (PSQ) corresponds with functionality and efficiency. PSQ is delineated as the users' perception of the system performance based on their experience[8]. Perceiving the system quality through technical and functional mechanisms elaborates the intention to use[14]. HCI literature on quality characteristics and behavior towards technology adoption pertain profound studies in various field of life. Quality features in IoT services positively influenced the behavioral intention to accept smart mobility services in Malaysia [12]. With the TAM model, quality features positively influence the attitude and behavior of end-user [15], and managing the privacy and efficacy aspects possesses the vital implication in the adoption process [16]. While in the facial recognition system, system quality features from the user point of view on the way to use will deliberate the willingness behavior. Hence, there is a proposition that perceived system quality (PSQ) can affect the behavioral intention to use the biometric facial recognition system.

Societal inspiration to encompass or refrain from performing certain actions is a matter of fact in today's world of digital integration. Technology use has witnessed the impact of social influence on user behavior[17]. Social influence or subjective norm is generalized as the users' perception that he/she should behave according to society's approval or disapproval. Social recommendation poses vital prominence in novel usage behavior. In the HCI scenario, people are more concerned about social approval and sometimes social influence or social norms acted as a stronger predictor of users' behavior[18]. In biometric devices acceptance, social influence had worked as a vibrant prognosticator in TAM and UTAUT model [7]. Consequently, views or perceptions of peers, neighbors, social community elements could be an obvious predictor towards installing the facial recognition system at homes.

By taking the consideration from previous studies and the HCI literature, we proposed the four hypothesize pertaining to the relationship between the TAM variables (i.e., PU and PEOU), perceived system quality (PSQ), and social influence (SI) towards acceptance of biometric facial recognition systems. These hypotheses are portrayed in Figure. 1 which were formed to test the surveyed data through structural equation modeling (SEM) and artificial neural network (ANN).

**H1:** PU will positively influence the behavioral intention (INT) to use the biometric facial recognition system.
**H2:** PEOU will positively influence the behavioral intention (INT) to use the biometric facial recognition system.
**H3:** PSQ will positively influence the behavioral intention (INT) to use the biometric facial recognition system.
**H4:** SI will positively influence the behavioral intention (INT) to use the biometric facial recognition system.

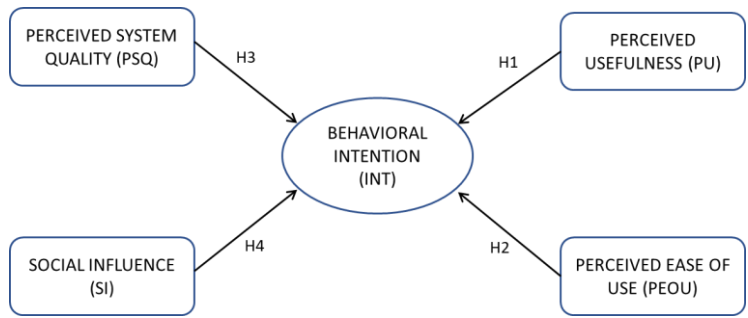

**Fig. 1.** Conceptual Framework

## 3. Methodology

To predict the users' behavior calls for a positivist approach which involves the cause and effect type study setting. The causation studies are conducted through the quantitative method as practiced in this study. In HCI research, there are numerously validated and approved theories to deduce the theoretical base, similarly, this study adapted the deductive approach by taking the TAM model as base theory. The data collection method engaged the convenience sampling of potential users towards biometric facial recognition systems. A questionnaire was developed with a five-point Likert-scale. The contents of the questionnaire were based on two sections with demographic details of respondents in section-A and variable questions or instruments in section-B. These questionnaire instruments of the proposed model were adopted from previous studies as "perceived usefulness" and "perceived ease of use" were taken from [19], perceived system quality from [15], and social influence was taken from [18]. An online survey was conducted to collect the appropriate size of the sample to perform the SEM and ANN analysis. The survey resulted in 475 respondents' data from the internet who had adequate information about the biometric facial recognition system for home security purposes such as main gate entrance or door lock systems based on face recognition mechanism. Their responses were coded as "strongly disagree=1" to "strongly agree=5" against the respective questionnaire instruments. The collected data then analyzed in computer programs IBM SPSS Statistics v24 and AMOS v23 for SEM (i.e., measurement and structural model) and ANN (i.e., sensitivity analysis and Root Mean Square of Errors-RMSE of Neural models).

## 4    Data Analysis

Analyzing the data in HCI research entails establishing the system to develop an accurate prediction scenario. This study used a hybrid analysis technique to best predict the antecedents towards behavioral intention. Section-A of the questionnaire represented the demographic results. It consisted of gender, age, and education categories. The resulted data showed the male with 81% and female 19% responses. The age groups of respondents were classified on generation bases as Gen-Z i.e., 5-25 years with 32% respondents, Gen-Y i.e., 26-40 years with 47% respondents, Gen-X i.e., 41-55 years with 17% respondents, and baby boomers i.e., 56-76 years with 4% respondents. While education qualification was labeled with 3 categories as Masters and Higher Degree with 31% respondents, Bachelor Degree holder respondents were 62% and less than Bachelor degree respondents were 7%. The demographic records depicted the respondents with a higher level of literacy rate along with younger and innovative mindset. The quantitative analysis for this research involved the Structural Equation Modeling (SEM) and Artificial Neural Network (ANN).

### 4.1 Structural Equation Modeling (SEM)

SEM analysis comprises a measurement model and a structural model. In the measurement model, the data reliability, validity, correlation, outer loadings, and model fit indices are being tested. In the structural model, the regression among model variables is calculated with path analysis[20]. SEM analysis requires a minimum of 200 samples to produce reliable results [21]. In our study, the larger sample size i.e., 475 respondents fortified the notion of better prediction results.

Reliability analysis, which describes the internal consistency of questionnaire items, measures through Cronbach's Alpha's recommended level (i.e., value more than 0.70) [17]. Data Validity represents a correlated relationship among the variable items through Composite Reliability (CR) value higher than 0.7, Average Variance Extracted (AVE) value higher than 0.50 and the correlation level among the variables should less than 0.90 [20], [22]. As per the criteria, reliability and validity results fulfilled all the parameters as shown in Table 1. Following reliability and validity analyses, the strength of questionnaire items or instruments was assessed through outer-loadings. These are the correlation between items and their corresponding variables with a recommended value of higher than 0.50 [9]. As per Table 2, all the outer loadings of proposed model items met the benchmarked value that ranged from 0.582 to 0.909.

To proceed for the final step of SEM analysis i.e., path analysis or hypothesis testing, model fit indices were tested. These fit indices indicate the goodness level of the proposed model to run the path analysis. For this study, four fit indices i.e., CMIN/DF (recommended value from 1 to 3), RMSEA (recommended value less than 0.06), SRMR (recommended value less than 0.08), and PClose (recommended value more than 0.05) were tested in AMOS v23 [22]. The analysis showed CMIN/DF = 2.104, RMSEA = 0.048, SRMR = 0.052 and PClose = 0.688 by achieving the goodness of fit index to conduct the path analysis. The final phase of SEM analysis reiterates the hypothesis testing by assessing the regression weights, critical ratio, or t-statistics (>1.96) and significance level (<0.05) to portray the hypotheses status as supported or not. In the SEM framework, as illustrated in Figure 2, the R-Squared value of 0.63 showed the 63% positive change due to the PU, PEOU, PSQ, and SI towards behavioral intention (INT) to use the biometric facial recognition system. In Table 3, hypothesis testing results demonstrated that all proposed hypotheses supported the conceptual framework by meeting the recommended values of critical ratio (i.e., CR>1.96) and p-values (i.e., p<0.05) [23]. In the resulted model, Hypothesis 3 explained the positive effective relationship between perceived system quality and behavioral intention. It showed a higher value of estimate i.e., 0.66, which is a standardized regression coefficient, to explore the impact variance in the relationship. According to the SEM analysis, all the factors positively impacted the behavioral intention (INT) and PSQ emerged as the main predictor to use the biometric facial recognition system.

**Table 1.** Reliability, Validity and Correlation Results

|  | Cronbach's Alpha | CR | AVE | SI | PEOU | PU | INT | PSQ |
|---|---|---|---|---|---|---|---|---|
| **SI** | 0.897 | 0.92 | 0.712 | 1 | | | | |
| **PEOU** | 0.927 | 0.94 | 0.775 | 0.18 | 1 | | | |
| **PU** | 0.893 | 0.92 | 0.701 | 0.37 | 0.340 | 1 | | |
| **INT** | 0.887 | 0.91 | 0.688 | 0.45 | 0.389 | 0.55 | 1 | |
| **PSQ** | 0.912 | 0.93 | 0.742 | 0.51 | 0.376 | 0.63 | 0.69 | 1 |

Table 2. Outer Loadings

| Variable | Items | Outer Loadings | Variable | Items | Outer Loadings |
|---|---|---|---|---|---|
| Social Influence | SI1 | 0.718 | Perceived Usefulness | PU4 | 0.893 |
|  | SI2 | 0.816 |  | PU5 | 0.883 |
|  | SI3 | 0.757 | Perceived System Quality | SQ1 | 0.757 |
|  | SI4 | 0.909 |  | SQ2 | 0.872 |
|  | SI5 | 0.759 |  | SQ3 | 0.757 |
| Perceived Ease of Use | PE1 | 0.582 |  | SQ4 | 0.889 |
|  | PE2 | 0.806 |  | SQ5 | 0.845 |
|  | PE3 | 0.855 | Behavioral Intention | INT1 | 0.813 |
|  | PE4 | 0.869 |  | INT2 | 0.794 |
|  | PE5 | 0.889 |  | INT3 | 0.753 |
| Perceived Usefulness | PU1 | 0.728 |  | INT4 | 0.742 |
|  | PU2 | 0.867 |  | INT5 | 0.801 |
|  | PU3 | 0.871 |  |  |  |

Table 3. Hypothesis Testing

| | Hypotheses | Estimates | S.E | Critical Ratio | P-Value | Result |
|---|---|---|---|---|---|---|
| H1 | PU→INT | .140 | .043 | 3.230 | 0.001 | Supported |
| H2 | PEOU→ INT | .129 | .053 | 2.436 | 0.015 | Supported |
| H3 | PSQ→ INT | .666 | .074 | 8.970 | 0.000 | Supported |
| H4 | SI→ INT | .212 | .078 | 2.710 | 0.007 | Supported |

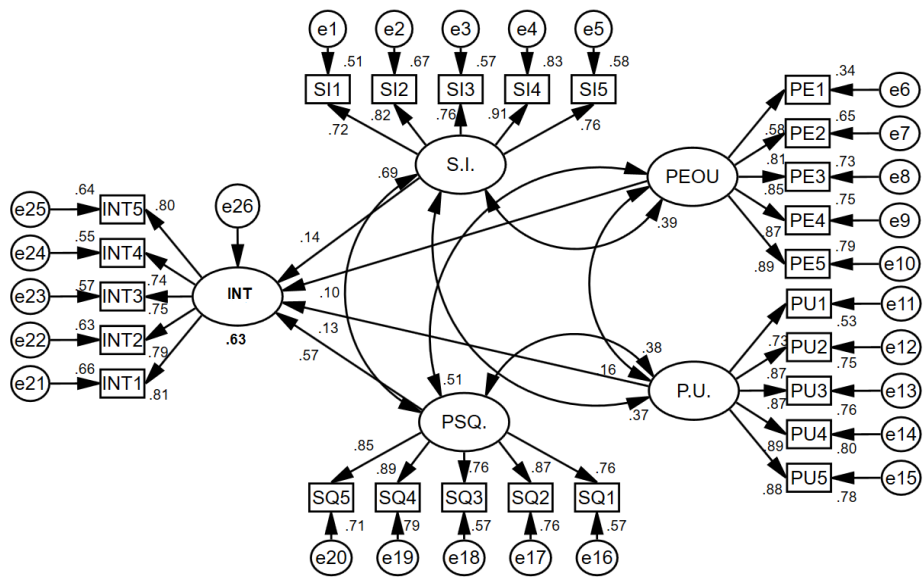

**Fig. 2.** Structural Equation Modeling (SEM) Framework

## 4.2 Artificial Neural Network (ANN)

The multi-analytical approach to predict the antecedent in the HCI study is comparatively topical in recent days. A simpler approach of data analysis comprises linear statistical analysis or SEM analysis that deals with linear relationships among model variables. As the collected data cannot be ideally linear to demonstrate the true assumption of multivariate data analysis and SEM analysis is unable to portray both linear and non-linear relationships in the tested model [24]. To orchestrate the non-linear relationship in the framework, the ANN method is used with SEM. ANN is dynamically useful in prediction accuracy. The ANN analysis doesn't require hypothesis testing and fulfilling the multivariate data assumption to predict the factors' strengths. Towards the hybrid analysis of ANN with SEM, the only variables of the SEM model that supported the hypotheses are included in the ANN model [25]. Therefore, in this study, four independent variables as the significant predictors towards INT after hypothesis testing were counted in the ANN model. The ANN model entails the input layer, output layer, and hidden layer. The input layer is based on the predictors or independent variables as the neuron or nodes that proved significant in SEM analysis. The output layer is the dependent variable of the model as INT in this study. The hidden layer counts from one to more depending on the intricacy of the framework. In HCI studies, one hidden layer is commonly used however, more than one hidden layer enhances the prediction accuracy[26]. The neural network model was analyzed in SPSS v24 through Multilayer Perceptron (MLP) tool via Feedforward-back Propagation (FFBP). The input layer was based on four predictors (i.e., PSQ, PI, SI, and PE) as the nodes of the layer, two hidden layers were used to surge the prediction accurateness as sketched in Figure 3. The partition of data was categorized as a 9:1 ratio in the model. As the more data is assigned to the training sample the better the model result will achieve, therefore 90% of data was allocated to train the sample and 10% data was assigned to test the sample. The architecture of the model was customized to two hidden layers with a Sigmoid function used as the activation function of neurons in both hidden and output layers [24]. Training of the data was based on Batch typology and the optimization algorithm was used as a "scaled conjugate gradient". To manage the over-fit problem, ten-fold cross-validation of the ANN model was conducted. The predictive accuracy of the network model is calculated by Root Mean Square of Errors termed as RMSE. It is also termed as model fit analysis for the ANN model. The value of RMSE was measured on all ten ANN models and then means values with standard-deviation were calculated, as illustrated in Table 4. The values of RMSE in the resulted analysis were quite low as the average RMSE of the Training network was 0.117 and the Testing network was 0.111 that portrayed the precise prediction of the ANN model [24], [27]. Following the ten ANN models prediction accuracy validation, the sensitivity analysis (i.e., importance and normalized importance) was conducted. This test described the prediction strength of independent variables (i.e., PU, PEOU, PSQ, and SI) towards the dependent variable (i.e., INT). The importance and normalized importance values of 10 models were calculated and the mean value of these sensitivity analyses explored that PSQ has the highest level of importance with 0.659 as in Table 5 and normalized importance with 100% as in Figure 4. PSQ led as the most significant predictor towards INT followed by the SI. ANN's Importance values are parallel to standardized regression weights in SEM as in both analyses the PSQ acted as significant regression elements in SEM and normalized important factor in ANN. According to the Neural analysis, perceived system quality is outlined as the most vibrant predictor towards the biometric facial recognition system use at homes.

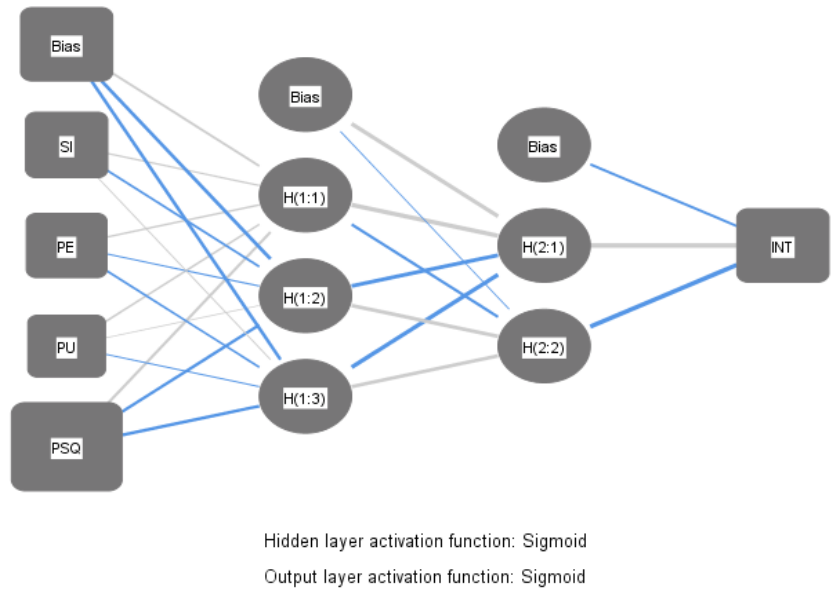

Hidden layer activation function: Sigmoid
Output layer activation function: Sigmoid

**Fig. 3.** Artificial Neural Network (ANN) Model

**Table 4.** RMSE of ANN Models

| Network | Training (90%) | Testing (10%) |
|---|---|---|
| ANN1 | 0.117 | 0.086 |
| ANN2 | 0.112 | 0.121 |
| ANN3 | 0.113 | 0.118 |
| ANN4 | 0.132 | 0.126 |
| ANN5 | 0.12 | 0.121 |
| ANN6 | 0.111 | 0.114 |
| ANN7 | 0.118 | 0.086 |
| ANN8 | 0.117 | 0.117 |
| ANN9 | 0.112 | 0.121 |
| ANN10 | 0.115 | 0.095 |
| RMSE Mean | 0.117 | 0.111 |
| RMSE SD | 0.006 | 0.015 |

**Table 5.** Average Normalized Importance

|  | SI | PEOU | PU | PSQ |
|---|---|---|---|---|
| **Importance** | 0.148 | 0.125 | 0.068 | 0.659 |
| **Normalized Importance** | 22.50% | 18.90% | 10.40% | 100.00% |

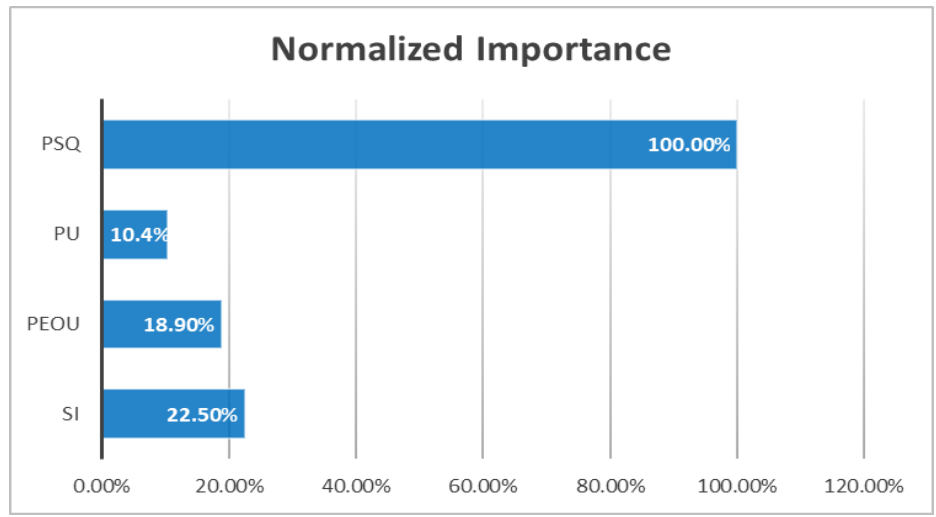

**Fig. 4.** Average Normalized Importance Chart

## 5. Conclusion

The study designated to apprehend the dynamic factors of behavioral intention towards the adoption of the biometric facial recognition system. The SEM result showed that each antecedent of conceptual framework i.e., PU, PEOU, PSQ, and SI had a significant relationship with INT. The significance level among independent variables vitalized by PSQ with a standardized regression estimate of 0.666 because the biometric facial recognition, being a novel digital tool, is mainly assessed through its functionality mechanism prior to adoption [15]. Usefulness and convenience to use of technology had a positive impact on adoption behavior. Societal approval also played a substantial role in behavioral modeling in this study. The prediction accuracy of the ANN model was achieved by the smaller RMSE value. In ANN Model, PSQ appeared as the most significant predictor with a 0.659 normalized score and 100% normalized importance towards the facial recognition system acceptance. The multimethod analysis reinstated that the perceived system quality in advanced digital tools is a matter of importance to the urban diffusion of smart home technologies.

Moreover, the majority of the respondents belonged to higher literacy clusters (i.e., more than 90% of respondents with Bachelor's and Master's level education), and representation of the millennial generation age group in sample size was around 50%. It shows the inclination towards the usability of the modern digital tool such as the biometric facial recognition system at homes is mainly backed by the millennials and higher education level. However, the perception of using such facial recognition system at public places will pose the privacy concern for the same age group [28] and future research should condense this dimension of home versus outdoor technology acceptance.

The study is a novel contribution to the literature of human-centric computing by exploring the behavioral elements through artificial neural network (ANN) analysis combining with structural equation modeling (SEM). The study will support the stakeholders to understand the usability scenarios of the biometric facial recognition system for home security purposes. Research on non-contact sensor technology like biometric facial recognition entails the trend towards digital transformation in personal life that eventually diffuses to professional life and community state. The study input to HCI sciences postulates that the interconnection of system features such as quality aspects towards usability has more inclined importance than the system's usefulness, ease of use, and societal influence. The functional and technical features of the facial recognition system should comply with security standards such as mitigating the false positive detection in light variations, face color disparities and providing the two-way authentication. The research is the primary initiative in biometric facial recognition literature from users centered pace that would instigate the researchers and academicians to dwell in more insight studies and analyses. The study will support the managers to understand the importance of PSQ in the implementation phase of smart technology. This research is also a new methodological participation in the HCI field with multi-analytical practice in data analysis towards user behavior modeling in the smart home scenario. The empirical findings also proved the robustness of the TAM model in the HCI domain with SEM-Neural analysis as complying with the literature [29]. In ANN modeling, using more than one hidden layer towards prediction accuracy has added the new approach in technology acceptance studies.

The study strived to summarize the fundamental elements of behavioral modeling in the facial recognition system through a cross-sectional study. Future research is encouraged to conduct a longitudinal analysis during technology deployment phases such as from pre-implementation to integration stage and post-implementation analysis. It would be diverged from initial analysis results and would provide the different scenarios of users' acceptance understandings. Furthermore, this research used a limited number of variables in behavioral assessment, while using personality traits such as personal innovativeness, self-efficacy, digital competence, etc. could drive the HCI scenario more profoundly. Adding more aspects of systems in future studies like privacy concerns, security risk, and intrusiveness apprehensions would enhance the human-centric computing dimensions to elaborate the system integration. The study was conducted to anticipate the home user's behavior for facial recognition, by using the SEM-Neural analysis, future studies can investigate the employee's behavior towards this technology at the workplace. This study was limited to use a convenience sampling method in data collection which was related to non-generalizability and bias factor while probability sampling could generalize the results and inferences. Due to the nature of technology and its scarce dispersion and integration around the global society, the study lacked the inclusion of geographical aspects in data collection which enhanced the analysis limitation in perceiving the location-based technology behavior. Future researches can overcome such elements by adding geographical representation in responses.